\documentclass[useAMS,usenatbib,Letter]{mn2e}
\usepackage{graphicx}
\usepackage[dvips]{color}
\usepackage{txfonts}
\usepackage{amssymb}
\usepackage{natbib}

\title[K2's first view of O-star variability]{Kepler's first view of
  O-star variability: K2 data of five O stars in Campaign\,0 as a proof-of-concept
  for O-star asteroseismology\thanks{Based on the data gathered with NASA's
Discovery mission
    {\it Kepler} and with the  {\sc HERMES} spectrograph, installed at the Mercator
    Telescope, operated on the island of La Palma by the Flemish Community, at
    the Spanish Observatorio del Roque de los Muchachos of the Instituto de
    Astrof\'{\i}sica de Canarias and supported by the Fund for Scientific
    Research of Flanders (FWO), Belgium, the Research Council of KU\,Leuven,
    Belgium, the Fonds National de la Recherche Scientific (F.R.S.--FNRS),
    Belgium, the Royal Observatory of Belgium, the Observatoire de Gen\`eve,
    Switzerland and the Th\"uringer Landessternwarte Tautenburg, Germany.}}

\author[B.\ Buysschaert et al.]{B.\ Buysschaert$^{1,2}$\thanks{E-mail:
    bram.buysschaert@obspm.fr}, C.\ Aerts$^{2,3}$, S.\ Bloemen$^{3}$, J.\
  Debosscher$^2$, C.\ Neiner$^1$, 
M.\ Briquet$^{4,1}$\thanks{F.R.S.-FNRS Postdoctoral Researcher, Belgium},
\and J.\ Vos$^2$, 
P.\ I.\ P\'apics$^2$\thanks{Postdoctoral Fellow of
    the Fund for Scientific Research of Flanders (FWO), Belgium},
R.\ Manick$^2$, V.\ Schmid$^2$\thanks{Aspirant
    PhD Student of the Fund for Scientific Research of Flanders (FWO), Belgium},
  H.\ Van Winckel$^2$, A.\ Tkachenko$^2$\thanks{Postdoctoral Fellow of
    the Fund for Scientific Research of Flanders (FWO), Belgium}\\
  $^1$ LESIA, Observatoire de Paris, PSL Research University, CNRS, Sorbonne
  Universit\'es, UPMC Univ.\ Paris 06, Univ.\ Paris Diderot, Sorbonne Paris
  Cit\'e, France\\
  $^{2}$Instituut voor Sterrenkunde, KU\,Leuven, Celestijnenlaan 200D, 3001
  Leuven, Belgium\\
  $^{3}$Department of Astrophysics/IMAPP, Radboud University Nijmegen, 6500 GL
  Nijmegen, The Netherlands\\
  $^4$  Institut d'Astrophysique et de G\'eophysique, Universit\'e de Li\`ege, Quartier
Agora, Allée du 6 ao\^ut 19C, B-4000 Li\`ege, Belgium
}

\begin{document}

\date{Accepted ?; Received  2015 June 17; in original form 2015 June 17}

\pagerange{\pageref{firstpage}--\pageref{lastpage}} \pubyear{2015}

\maketitle

\label{firstpage}

\begin{abstract}
  We present high-precision photometric light curves of five O-type stars
  observed with the refurbished {\it Kepler\/} satellite during its Campaign\,0.
  For one of the stars, we also assembled high-resolution ground-based
  spectroscopy with the {\sc hermes} spectrograph attached to the 1.2-m Mercator
  telescope.  The stars EPIC\,202060097 (O9.5V) and EPIC\,202060098 (O7V)
  exhibit monoperiodic variability due to rotational modulation with an
  amplitude of 5.6\,mmag and 9.3\,mmag and a rotation period of 2.63\,d and
  5.03\,d, respectively.  EPIC\,202060091 (O9V) and EPIC\,202060093 (O9V:pe)
  reveal variability at low frequency but the cause is unclear.  EPIC\,202060092
  (O9V:p) is discovered to be a spectroscopic binary with at least one
  multiperiodic $\beta\,$Cep-type pulsator whose detected mode frequencies occur
  in the range $[0.11,6.99]$\,d$^{-1}$ and have amplitudes between 0.8 and
  2.0\,mmag.  Its pulsation spectrum is shown to be fully compatible with the
  ones predicted by core-hydrogen burning O-star models.  Despite the short
  duration of some 33\,d and the limited data quality with a precision near
  100\,$\mu$mag of these first K2 data, the diversity of possible causes for
  O-star variability already revealed from campaigns of similar duration by the
  MOST and CoRoT satellites is confirmed with {\it Kepler}. We provide an
  overview of O-star space photometry and give arguments why future K2
  monitoring during Campaigns\,11 and 13 at short cadence, accompanied by
  time-resolved high-precision high-resolution spectroscopy opens up the
  possibility of in-depth O-star seismology.
\end{abstract}

\begin{keywords}
  Asteroseismology -- Stars: massive -- Stars: rotation --
  Stars: oscillations (including pulsations) -- Techniques: photometry -- Techniques: spectroscopy
\end{keywords}

\section{Introduction}

Recent high-precision uninterrupted high-cadence space photometry implied a
revolution in the observational evaluation of stellar structure models for
various types of low-mass stars, covering almost their entire evolutionary
path. Many results were obtained for {\it sun-like stars and red giants\/}
undergoing solar-like oscillations excited
stochastically in their convective envelope, thousands of which were monitored.  The
interpretation of their oscillation spectrum is readily achieved by relying on
existing methodology developed for helioseismology, when
extended with the interpretation of gravity-dominated dipole mixed modes
\citep[e.g.,][]{Bedding2011,Mosser2012,ChaplinMiglio2013}.  Although many open
questions remain on the evolutionary state of blue horizontal branch stars
\citep[e.g.,][]{Oestensen2012}, progress was also achieved from 
asteroseismology of a few subdwarf B stars and white dwarfs.  These are {\it evolved
low-mass stars}, the former of which have lost almost their entire hydrogen
envelope while passing through the helium flash in a binary configuration
\citep[e.g.,][]{VanGrootel2010,Charpinet2011}. The latter are the {\it compact
remnants of low-mass stars\/} that were very successfully studied from ground-based
mmag-precision asteroseismology more than two decades ago \citep[e.g.,][and
references therein]{Corsico2008} but they were hardly monitored seismically from
space due to their faintness and fast oscillations of tens of seconds
\citep{Oestensen2011,Kim2011,Hermes2011}.  The heat-driven pressure and/or
gravity modes detected in hundreds of {\it intermediate-mass stars\/} are much harder to
analyse and interpret theoretically than the stochastically-excited solar-like
pulsators, but major progress is also underway for such objects
\citep{Kurtz2014,Saio2015,VanReeth2015}.

From the point of view of chemical evolution of our Milky Way, {\it
  supernova progenitors\/}, i.e., stars with masses above $\sim\,9\,$M$_\odot$
having an extended convective core and a radiative envelope at birth, are the
objects that matter \citep{Maeder2009}.  Despite their large importance for
massive star evolution theory, {\it O stars have hardly been monitored in
  high-cadence uninterrupted space photometry}.  If available, such monitoring
was so far restricted to mmag precision obtained by the WIRE, MOST, and Coriolis
(SMEI instrument) satellites, while CoRoT observed six O stars during 34\,d
leading to a level of brightness variations with $\sim$100\,$\mu$mag in
precision.  Except for the SMEI light curves, which are seven to eight months
long, all of the available time series have a duration of the order of a month only.

Unfortunately, O stars were absent in the nominal {\it Kepler\/} Field-of-View (FoV),
hence we do not have long-term $\mu$mag-level precision photometry of O stars at
hand, preventing detailed asteroseismic calibrations of massive star models as
it was achieved for the less massive B stars \citep[e.g.][for a
summary]{Aerts2015}. Similarly to the O stars, B stars have an extensive
convective core but they do not exhibit a strong radiation-driven wind.  The two
B stars that were seismically modelled based upon period spacings of their
high-order dipole gravity-mode oscillations are ultra-slow rotators ($v\sin
i<10\,$km\,s$^{-1}$). Their modelling required the inclusion of extra diffusive
mixing, in addition to core overshooting, to bring the theoretical models in
agreement with the $\mu$mag-precision seismic data of duration five months
\citep[for the B3V star HD\,50230,][]{Degroote2010} and four years \citep[for the
B8V star KIC\,10526294,][]{Moravveji2015}.  It would be highly beneficial to
perform similar seismic inference studies for O-type stars, whose theoretical
models are the most uncertain of all the mass ranges, while they are mainly
responsible for the global chemical enrichment of the Milky Way.

Two rapidly rotating O stars were monitored by the MOST satellite. It concerns
$\xi\,$Per (O7IIInf) with $v\sin i\simeq 200\,$km\,s$^{-1}$ revealing
variability due to rotational modulation with a rotation period of 4.18\,d, in
the absence of oscillations \citep{Ramiaramanantsoa2014} and $\zeta\,$Oph
(O9.5Vnn, $v\sin i\simeq 400\,$km\,s$^{-1}$) whose WIRE, MOST, SMEI photometry
led to the detection of multiperiodic variability due to oscillations with
frequencies in the range $[1,10]$\,d$^{-1}$ and amplitudes up to $\simeq
10\,$mmag. These modes turn out to be variable on a time scale of some 100\,d
\citep{Howarth2014}. The rapid rotator $\zeta\,$Pup was monitored by SMEI during
several long-term runs, but these data revealed only two dominant frequencies,
whose cause and identification remain unclear; they might be connected with
non-adiabatic gravity modes \citep{HowarthStevens2014}. The variability derived
from the space photometry of these three rapid rotators is in line with
assessments from their high-precision time-resolved line-profile variability. It
is also in agreement with high-precision time-resolved spectroscopy of the
rapidly rotating O9Vp star HD\,93521, whose multiperiodicity has a yet unclear
origin \citep[e.g.,][$v\sin i\simeq 400\,$km\,s$^{-1}$]{Fullerton1996,Rauw2008}.
Seismic modelling of rapidly rotating O stars was not achieved so far.

One of the CoRoT short runs (SRa02) was devoted to the monitoring of six O-type
stars during 34\,d in the asteroseismology CCD of the mission, having a cadence
of 32\,s \citep{Auvergne2009}. Among these six stars were two binaries.
Plaskett's star (HD\,47129) is a high-mass interacting binary (O7.5I+O6II) with
a magnetic secondary \citep{Grunhut2013}. Its CoRoT light curve revealed
hints of gravity-mode oscillations with frequencies near $\sim$0.8\,d$^{-1}$
and harmonics, as well as rotational modulation with multiples of the orbital
period of 14.39625\,d \citep{Linder2008} --- see \citet{Mahy2011} for a
discussion. A seismic interpretation was not possible for this complex
system. The eccentric ($e=0.59\pm0.02$) long-orbit ($P_{\rm orbit}=829\pm4$\,d)
binary HD\,46149 revealed a primary with stochastic pressure-mode oscillations
with frequencies in the range 3.0 to 7.2\,d$^{-1}$. These correspond to a
regular frequency pattern of separation 0.48\,d$^{-1}$. Unfortunately, these p
modes did not lead to an unambiguous interpretation in terms of seismic models
\citep{Degroote2010}. Further, the ``red noise'' power excess found by
\citet{Blomme2011} in the CoRoT amplitude spectra of the three moderately
rotating O stars HD\,46150 (O5.5f, $v\sin i\simeq 100\,$km\,s$^{-1}$), HD\,46223
(O5f, $v\sin i\simeq 100\,$km\,s$^{-1}$), and HD\,46966 (O8.5V, $v\sin i\simeq
50\,$km\,s$^{-1}$) were recently interpreted in terms of convectively-driven
internal gravity waves \citep{AertsRogers2015}.

The only seismic inference was
achieved for the  O9V star HD\,46202, which revealed
$\beta\,$Cep-type pressure-mode oscillations with frequencies in the range from
0.5 to 4.9\,d$^{-1}$ \citep{Briquet2011}. Seismic modelling of this star led to a mass
$M=24.1\pm0.8$\,M$_\odot$, an age of $4.3\pm0.5$\,Myr, and a core overshooting
parameter of $0.10\pm0.05$ expressed in units of the local pressure scale height
and using a step-function formulation.

To remedy the lack of seismic calibration of massive star models, we defined a
K2 \citep{Howell2014} observing programme of O-type stars with the aim to
monitor a sample of such objects at $\mu$mag precision during 3-month runs. The
present  work reports on the first results of this programme and offers a
proof-of-concept of K2's capabilities for future O-star asteroseismology.

\begin{figure*}
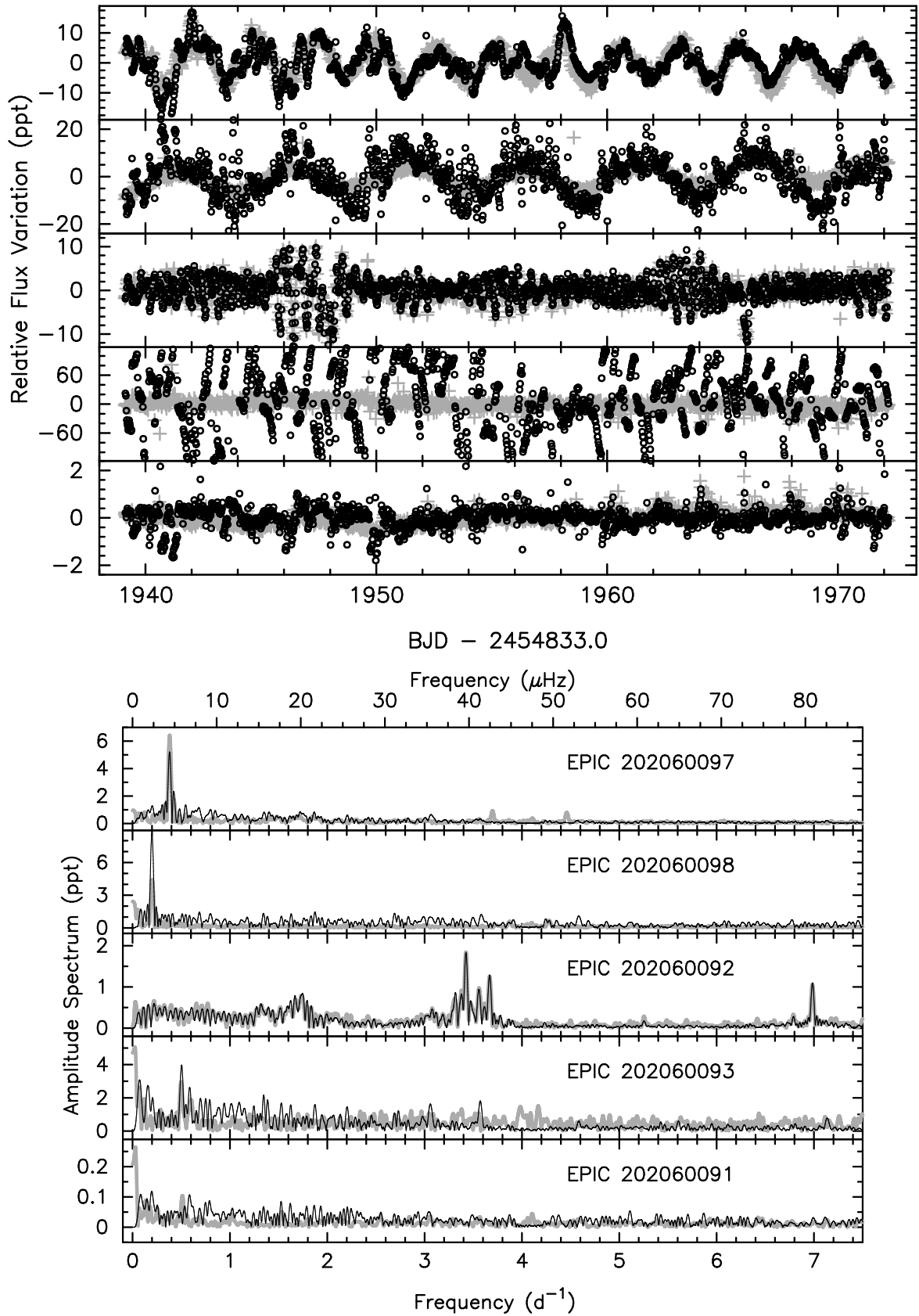

\begin{center}
\rotatebox{270}{\resizebox{11.cm}{!}{\includegraphics{LC.eps}}}\\[8pt]
\rotatebox{270}{\resizebox{11.cm}{!}{\includegraphics{FT.eps}}}
\end{center}
\caption{K2 light curves (top) and their amplitude spectra (bottom) of the
  five monitored O stars.  The black circles and thin black lines are derived from our reduction of
  the raw pixel data, while the grey crosses and thick grey lines are the
  \citet{Vanderburg2014} light curves. }
\label{LC}
\end{figure*}

\section{Extraction of the Light Curves}

\begin{table*}
\caption{The five O stars observed in Campaign\,0 of K2, with some information
  from the online SIMBAD database and the EPIC catalogue. The dominant
  frequency in the K2 data is listed, along with its amplitude.}
\centering
\tabcolsep=8pt
\begin{tabular}{lcccclll}
\hline
\hline
EPIC ID & Other ID & RA (J2000) & DEC (J2000) & V (mag) & SpT & Frequency (d$^{-1}$) & Amplitude (ppt)\\
\hline
202060091	& HD 44597			& 06 23 28.538	& +20 23 31.63
& 9.05	& O9V  & ---  & ---\\
202060092	& HD 256035			& 06 22 58.243	& +22 51 46.15
& 9.21	& O9V:p & 3.4248 & 1.827\\
202060093	& HD 255055			& 06 19 41.647	& +23 17 20.22
& 9.40	& O9V:p(e) & --- & --- \\
202060097	& Cl* NGC 2175 H 98	& 06 08 31.999	& +20 39 24.01	& 13.7
& O9.5V & 0.3801 & 5.209\\
202060098	& 2MASS J06245986+2649194	& 06 24 59.870	& +26 49 19.42
& 15.11	& O7V & 0.1988 & 8.594 \\
\hline
\end{tabular}
\label{targets}
\end{table*}
We proposed the five O-type stars listed in Table\,\ref{targets} for photometric
monitoring with K2 in Campaign\,0.  We developed a photometric reduction method
that accounts for the spacecraft (re)pointing and the non-uniform pixel response
of the K2 CCDs, starting from the pixel frames in the MAST archive (release of 7
Nov.\ 2014).  The time series of Campaign\,0 contain two large gaps related to
safety events of the spacecraft. We only used data after the largest safety
event, i.e., after the {\it Kepler\/} Barycentric Julian Day 1939.1, providing us with
photometric measurements covering 33.038\,d. These data have a frequency
resolution corresponding to the Rayleigh limit of
$1/33.038$\,d$=0.0302$\,d$^{-1}$ (0.35$\mu$Hz), which represents an upper limit
for the frequency uncertainty of any periodic signal in the data
\citep[][Chapter\,5]{Aerts2010}. Further, we ignored all measurements taken
during a thruster firing or a repointing event of the spacecraft.

With the aim to exclude as much as possible instrumental effects in the light
curves, we first defined an optimal constant mask {\it customized to each of the
  five targets from visual inspection}, such that blending with nearby targets
is avoided as much as possible. We show these masks in the left panels of
Figs\,\ref{91-mask} to \ref{98-mask}.  The counts for each pixel in these masks
were summed up to provide the time series of the raw photometric
measurements. To account for the varying background during the measurements, we
used iterative sigma clipping.  All pixels of the frame at a given time step
were deselected when their signal was larger than 1.2 times the median signal of
the frame. This procedure was iterated until we found a stable median for that
time step and no further deselection was needed.  The achieved median was then
taken to be the background signal in a given pixel in the frame.  We subtracted
this background level from the raw light curve, accounting for the considered
number of pixels in the mask.  

Outliers were identified and removed by self-flat-fielding through spline
fitting, along with further detrending of the photometry by third-order
polynomial fitting, in an iterative scheme.  In each step, the photometry was
divided by the polynomial fit, which allowed us to work with relative brightness
units instead of electrons per second. Five iteration steps turned out to be
sufficient.  

This raw light curve was then used as a starting point to account for the
instrumental effects caused by K2's attitude control system, which involves a
roll manoeuvre every six hours. Due to that, we had to redetermine the position
of the star on the CCD and connect it to the instrumental brightness variations
that occurred at that position. In order to derive an appropriate relation for
the correction factor versus star position, we determined the position of the
star on the CCD by fitting a 2D Gaussian function to the photometry,
representing the point-spread-function.  Following
\citet{VanderburgJohnson2014}, we rotated that 2D position according to its
largest eigenvalue and fitted a fifth-order polynomial, allowing us to define an
arclength connected with the instrumental brightness for each time step. We then
assumed that all the power in the Fourier transform of the time series of the
measured brightness centred at
4.08\,d$^{-1}$ and its higher harmonics at 8.16, 12.24, 16.32, and
20.40\,d$^{-1}$ had instrumental origin.  A bandpass filter in the Fourier
domain was constructed using the bands $j\times4.08\,$d$^{-1}\pm 5\,f_{\rm
  res}$, with $f_{\rm res}$ the Rayleigh limit and $j=1,\ldots,5$.  All the
power in these bands was assumed to be due to instrumental effects, while all
power outside these bands was considered as stellar signal.  The inverse Fourier
transform then allowed to connect the instrumental brightness variations to the
position of the star on the CCD.  This method worked well for four of the five
stars, EPIC\,202060093 suffering from a too small onboard mask and saturation of
many pixels leading to leakage outside of the frame. The light curves are shown
with black circles in the upper panel of Fig.\,\ref{LC}. These flux variations
are expressed in parts-per-thousand (ppt).\footnote{The conversion factor
  between brightness variations expressed in mmag and flux variations expressed
  in ppt amounts to $2.5\log_{\rm 10}e=1.0857$}

In order to compare our reduction scheme for the individual stars and its
effectiveness, we also used the reduced light curves as made available by
\citet{Vanderburg2014}, release date: 10 April 2015.\footnote{{\tt
    https://www.cfa.harvard.edu/$\sim$avanderb/}} This is an updated reduction
based on their self-flatfielding approach combined with high-pass filtering,
designed from the viewpoint of optimal exoplanet hunting, as previously
developed by \citet{VanderburgJohnson2014} to which we refer for details.  While
our mask determination and noise treatment was done manually and tuned to each
star individually and locally, \citet{Vanderburg2014} used a semi-automated mask
determination and made use of the instrumental effects of thousands of stars.
Our method is tuned to get the optimal signal, while their method is focused on
reducing the global noise as much as possible. The masks used by Vanderburg are
shown in the right panels of Figs\,\ref{91-mask} to \ref{98-mask}.

Despite the quite different masking, the two independent reduction methods give
consistent results in terms of {\it frequencies\/}, the major difference
occurring in the lowest frequency regime of the Fourier transform. This is
similar to the difference already encountered for light curves of the nominal
{\it Kepler\/} mission when reduced globally from the viewpoint of optimal
planet hunting versus individually and tuned towards optimal asteroseismic
applications. For the latter, it is typically advantageous to consider more
pixels than done in the standard exoplanet pipeline, because the aim is to
gather as much signal as possible for a whole range of frequencies, given the
multiperiodic character of the variability. For exoplanet hunting, however, one
aims to find singly periodic low-amplitude signal and therefore avoids too noisy
pixels. Masking for asteroseismology as we have done here avoids complicated
detrending and filters out long-term instrumental frequencies more easily, as
illustrated and discussed in \citet{Tkachenko2013}, but is not optimal in terms
of global noise properties.

A comparison between the light curves resulting from the two independent
reduction methods for the five O stars considered here is given in
Fig.\,\ref{LC}.  In general, the agreement is good, except for EPIC\,202060098
and EPIC\,202060093. For the former star, the dominant frequency is fully
consistent but its amplitude differs in value, due to the different shape and
position of the chosen masks in the two methods. With our mask, we have avoided
the additional bright target contributing to the overall light curve. This
explains the higher amplitude compared to the ``Vanderburg'' light curve, while
the latter leads to less power at low frequencies (cf.\ bottom panel of
Fig.\,\ref{LC}). For EPIC\,202060093, both light curves suffer from the
saturation and leakage across the CCD, but again the frequency amplitude is
higher when derived from our light curve.  In the rest of the paper, we used our
reduced light curves to derive the frequencies and amplitudes of the five target
stars, after having verified that all the listed frequencies are consistently
recovered from those two versions.

\section{Frequency Analysis of K2 data}
\begin{table}
  \caption{The frequencies and amplitudes of EPIC\,202060092 determined from
    Fourier analysis followed by non-linear least-squares fitting. The computation
    of the errors is explained in the text. 
    The signal-to-noise
    ratio (SNR) was computed as the amplitude divided by the noise level of the
    residual light curve averaged over $[0,10]$\,d$^{-1}$.}
\centering
\tabcolsep=4pt
\begin{tabular}{lccrc}
\hline
\hline
ID & Frequency (d$^{-1}$) & Amplitude (ppt) & SNR & Remark\\
\hline
$f_{\rm 1}$&$3.4243\pm0.0054$&$1.827\pm 0.079$&$16.9$&\\
$f_{\rm 2}$&$3.6663\pm0.0066$&$1.352\pm 0.084$&$12.5$&\\
$f_{\rm 3}$&$6.9853\pm0.0088$&$1.091\pm 0.104$&$10.1$&\\
$f_{\rm 4}$&$3.3173\pm0.0113$&$0.938\pm 0.103$&$8.7$&\\
$f_{\rm 5}$&$3.5583\pm0.0089$&$0.998\pm 0.105$&$9.2$&\\
$f_{\rm 6}$&$1.7453\pm0.0092$&$0.941\pm 0.083$&$8.7$&$f_{3}/4$\\
$f_{\rm 7}$&$1.6693\pm0.0085$&$0.953\pm 0.106$&$8.8$ & $f_{4}/2$\\
$f_{\rm 8}$&$1.7193\pm0.0091$&$0.784\pm 0.106$&$7.3$ & $f_{1}/2$\\
$f_{\rm 9}$&$3.4863\pm0.0113$&$0.646\pm 0.096$&$6.0$ & $f_{3}/2$\\
$f_{\rm 10}$&$0.2223\pm0.0113$&$0.705\pm 0.100$&$6.5$ & $2f_{12}$\\
$f_{\rm 11}$&$0.3313\pm0.0125$&$0.585\pm 0.104$&$5.4$ & $3f_{12}$\\
$f_{\rm 12}$&$0.1123\pm0.0129$&$0.566\pm 0.074$&$5.2$& $f_1-f_4=f_2-f_5$\\
$f_{\rm 13}$&$1.3223\pm0.0134$&$0.535\pm 0.095$&$5.0$&\\
\hline
\end{tabular}
\label{star92}
\end{table}
\begin{figure*}
\begin{center}
\rotatebox{270}{\resizebox{9.cm}{!}{\includegraphics{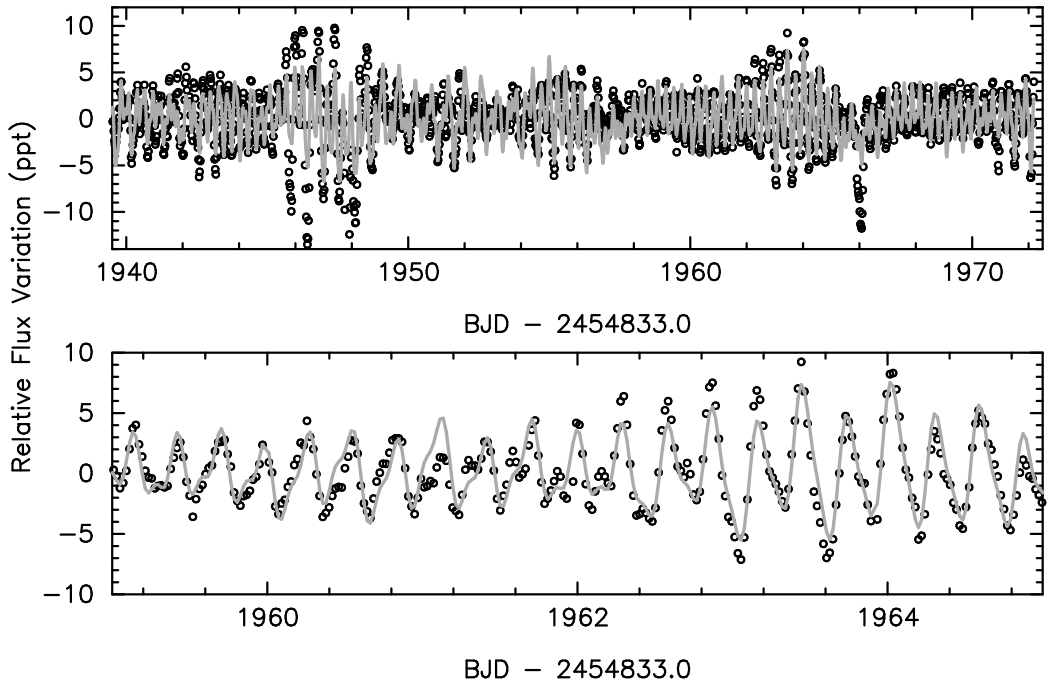}}}
\end{center}
\caption{Fit (grey line) to the K2 light curve of EPIC\,202060092 (black circles
  repeated from the central panel of Fig.\,\ref{LC}) based on the 13 frequencies
  listed in Table\,\ref{star92}.}
\label{92LCfit}
\end{figure*}

We computed the Lomb-Scargle periodograms of the two versions of the light
curves. These are shown in the lower panel of Fig.\,\ref{LC}. The thin black
lines are for our reduced light curves while the thick grey lines are for the
\citet{Vanderburg2014} light curves.
 
For the stars EPIC\,202060097 and 202060098, we find one isolated significant
frequency corresponding to a period of 2.631\,d and 5.030\,d, respectively
(Table\,\ref{targets}).  We interpret their variability as due to rotational
modulation.

The cause of the variability of EPIC\,202060091 and 202060093 is less
clear. Despite the limitations of the masks for EPIC\,202060093 (Fig.\,\ref{93-mask}), its
frequency spectrum is similar to the one of EPIC\,202060091, but there is a
factor more than ten difference in the level of variability. None of these two
stars shows clear periodic variability with isolated frequencies. Nevertheless,
they display several low-frequency peaks that stand out of the noise level
without being formally significant. This is somewhat similar to the frequency
spectra of three O stars observed with the CoRoT satellite, revealing red noise
power excess at low frequency that was recently
interpreted in terms of convectively driven internal gravity waves
\citep{AertsRogers2015}. However, we currently 
consider this interpretation as speculative.

EPIC\,202060092 turns out to be a multiperiodic pulsator compatible with
heat-driven oscillation modes, similar to HD\,46202 \citep{Briquet2011}.  We
performed iterative prewhitening and find 13 significant frequencies when we
adopt the conservative criterion of considering a frequency to be significant
when its amplitude reaches above five times the noise level in the residual
light curve averaged over the frequency range $[0,10]$\,d$^{-1}$. The
frequencies and amplitudes along with their errors are listed in
Table\,\ref{star92} while the fit to the light curve based on those 13
frequencies is shown in Fig.\,\ref{92LCfit}. For the computation of the 
frequency and amplitude errors, we took into account
that the formal errors resulting from a non-linear least squares fit are only in
agreement with those obtained in the Fourier domain in the case of uncorrelated
data with white noise and with sufficiently high frequency resolution
\citep[e.g.,][Chapter 5]{Aerts2010}.  Here, we encounter two complications: the
data have only a short time base of some 33\,d, leading to a limited resolving
power of $1.5/33.038\,$d = 0.04541\,d$^{-1}$ \citep{Loumos1978}. Further, as is
usually the case for highly sampled space photometry, the data may be
correlated. This requires incorporation of a correction factor in the error
estimates of the frequencies and their amplitudes
\citep{Schwarzenberg2003}. Following \citet{Degroote2009}, we computed this
correction factor to be 2.51 and obtained the error estimates listed in
Table\,\ref{star92}. The fit shown in Fig.\,\ref{92LCfit} is generally good, but
not perfect. This is due to there being additional frequencies whose amplitudes
are between 3 and 5 times the noise level. Due to the limited frequency
resolution, we are unable to pinpoint their value with respect to the
frequencies already listed in Table\,\ref{star92}
and this explains why a level of
unresolved beating still occurs in some parts of the light curve (bottom panel
of Fig.\,\ref{92LCfit}).

\section{Follow-up study of EPIC\,202060092}

The nature of the frequencies $f_6$ to $f_{13}$ listed in
Table\,\ref{star92} is unclear. Given the limited frequency precision,
they could either be identified as sub-multiples or combinations based on $f_1$
to $f_5$, or be due to independent g modes occurring in the densely-packed
gravity-mode frequency regime, or both.  The interpretation of all the detected
frequencies in terms of identification of their degree $\ell$, radial order $n$, and azimuthal
order $m$ requires additional observational information, such as the rotational
frequency of the star.  We also note that the light curve of EPIC\,202060092
shows a marked dip near day 1966 and remarkable beating from days 1946 to 1948
and to a lesser extent also from days 1963 to 1965 (Fig.\,\ref{92LCfit}, where
the listed and shown dates are with respect to {\it Kepler\/} Barycentric Julian
Data, indicated here as BJD\,2454833.0).  With the aim to
test if the dips  could be connected with binarity and to derive the projected
rotation velocity of the pulsator, we gathered spectroscopic measurements.

\subsection{EPIC\,202060092 is a spectroscopic binary}

\begin{figure}
\begin{center}
\rotatebox{0}{\resizebox{9cm}{!}{\includegraphics{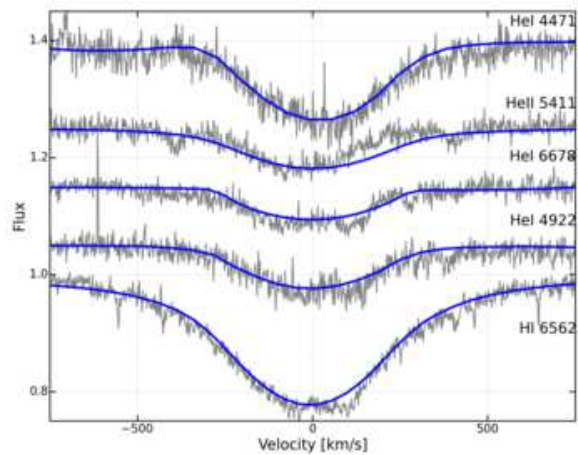}}}
\end{center}
\caption{Selected H and He lines in the highest SNR {\sc HERMES} spectrum 
of EPIC\,202060092. The blue line is the prediction for a NLTE atmosphere model
with parameters listed in the text.}
\label{92-lines}
\end{figure}

\begin{figure}
\begin{center}
\rotatebox{270}{\resizebox{6.5cm}{!}{\includegraphics{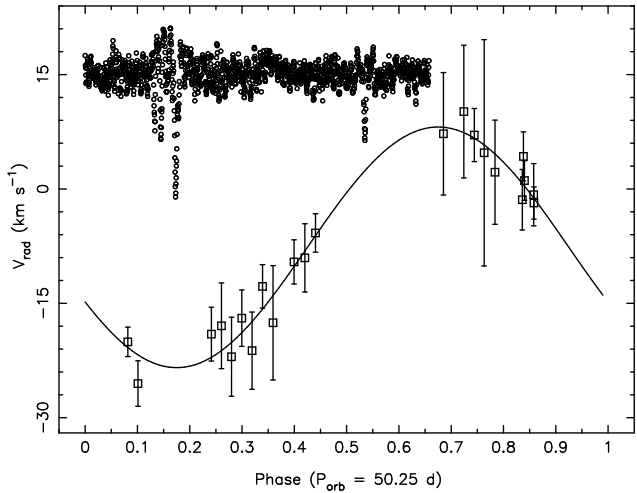}}}
\end{center}
\caption{Phased radial velocity measurements of EPIC\,202060092 deduced from its
  Balmer lines. The residual K2 light curve after prewhitening the fit in
  Fig.\,\protect\ref{star92} is indicated in the top part in ppt according to
  the same phase and shifted up with value 15 for visibility reasons.  }
\label{RVplot}
\end{figure}

The 1.2-m Mercator telescope is dedicated to the long-term monitoring of variable
stars, including heat-driven pulsators \citep[e.g.,][]{DeCat2007,Cuypers2009}
and evolved low-mass stars \citep[e.g.,][]{VanWinckel2009}. A large fraction of
the telescope time in the recent years has been used for the spectroscopic
monitoring of bright {\it Kepler\/} targets
\citep[e.g.,][]{Lehmann2011,Tkachenko2012,Beck2014,Niemczura2015,VanReeth2015}.
In view of its photometric behaviour and of recent theoretical predictions that
more than 70\% of all O-type stars occur in binaries \citep{Sana2012}, we added
EPIC\,202060092 to the observing programme for measurements with the {\sc
  hermes} spectrograph \citep{Raskin2011}.

A total of 22 high-resolution (HRF mode, resolution of 85\,000) spectra were
taken so far, with integration times between 600\,s and 1800\,s between November
2014 and March 2015.  The signal-to-noise ratio (SNR) per exposure ranged from
30 to 60.  The raw exposures were treated with the {\sc HERMES} pipeline,
including cosmic-hit removal, merging of orders, and barycentric
correction. Subsequently, all spectra were rectified following the method
outlined in \citet{Papics2013}.  

Comparison with rotationally broadened
synthetic spectra derived from plane-parallel NLTE O-star atmosphere models
\citep[][OSTAR2002 grid]{LanzHubeny2003} led to the estimates $T_{\rm
  eff}\simeq 35\,000\,$K, $\log\,g\simeq 4.5\,$dex,
$v\sin\,i\simeq 270\,$km\,s$^{-1}$ for solar metallicity and a fixed
microturbulent velocity of 10\,km\,s$^{-1}$. The agreement between the
theoretical line spectrum and the measured profiles for the highest SNR spectrum
is illustrated in Fig.\,\ref{92-lines} for a few selected lines. Despite the
modest SNR, it can be seen that structure occurs in some of the observed He
lines. This may be connected to the pulsations but higher SNR is needed to
firmly establish this.  

Given the scarcity of spectral lines due to the large broadening and limited
SNR, we estimated the radial velocity values from fits to the detected H and He
lines.  The outcome is very similar for the H$\alpha$, H$\beta$, H$\gamma$, and
H$\epsilon$ lines, so we averaged their radial velocity estimates and computed
the standard deviation as a proxy for the errors.  We detect radial velocity
variations with a peak-to-peak value around 30\,km\,s$^{-1}$. Given the low
amplitude of the photometric variations (cf.\ Table\,\ref{star92}) we do not
assign these radial-velocity variations to pulsations, but rather interpret them
in terms of binarity.  Globally, the velocities from the Balmer lines are
consistent with those based on the dominant He lines, such as HeI\,4922\AA,
HeI\,5875\AA, HeI\,6678\AA, HeII\,4541\AA, and HeII\,5411\AA, but for some of
those we encounter too large uncertainties to achieve conclusive values, as
already illustrated by Fig.\,\ref{92-lines}. This might point towards a
contribution of more than one star to some of these lines, but this needs
further study from new spectra with higher SNR.

Our current spectroscopy does not allow to deduce a high-precision orbital
solution yet.  Nevertheless, we clearly establish spectroscopic binarity and
find a rough preliminary period estimate of $P_{\rm orb}\sim 50\,$d from the
Balmer lines. These data were folded with the most likely period in
Fig.\,\ref{RVplot}, where we also show the phased residual K2 light curve after
prewhitening the 13 frequencies listed in Table\,\ref{star92} (i.e., the black
circles minus the grey line in Fig.\,\ref{92LCfit}).  We conclude that, with the
current estimate of the orbital period of the binary, we cannot associate all the
dips in the K2 photometry with eclipses and rather ascribe them to 
unresolved beating of pulsation modes, except perhaps the dip near day 1966. 
Large beating patterns as the ones seen near days 1946, 1955, and 1964
(Fig.\,\ref{92LCfit}) have indeed been detected before in multiperiodic massive
pulsators, both in high-cadence ground-based
\citep[e.g.][]{Handler2004,Handler2006} and space photometry
\citep[e.g.,][]{Papics2012}.  With the long-cadence sampling at 29\,min, the
unresolved nature of the light curve is not so surprising, because the fastest
pulsation mode is sampled only seven times per cycle. While such sampling works
fine for the analysis and interpretation of long-duration light curves of months
to years, there is a limitation to resolve beating patterns for a data set of
only 33\,d. The beating between frequencies $f_1$ and $f_4$, as well as the one
between $f_2$ and $f_5$, occurs with a period of 9.1\,d and the beating
phenomena we measure are separated by roughly this value, occurring near
days 1946, 1955, and  1964 (top panel of Fig.\,\ref{92LCfit}). The
nature of the isolated dip in the light curve at day 1966 remains unclear. That
decrease in brightness is not accompanied by local increases 
as one would expect from multiperiodic beating among linear pulsation modes and 
could therefore still be caused by binarity. Indeed,  according to Fig.\,\ref{RVplot}, the
dip occurs near orbital phase $\sim 0.5$ and, given the large uncertainty 
on the current estimate of the orbital period, could be caused by an eclipse.
Unfortunately, there are no {\sc Hipparcos data} data available to add to the K2
photometry for this star.

\subsection{First modelling attempts}

\begin{figure*}
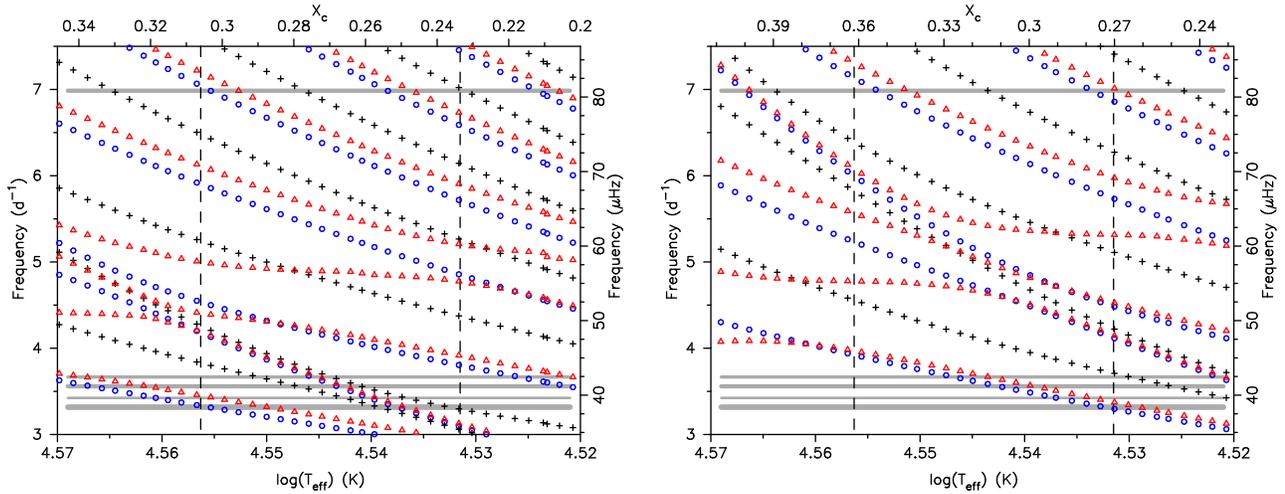

\begin{center}
\rotatebox{270}{\resizebox{6.5cm}{!}{\includegraphics{M35-modes.eps}}}\hspace{0.4cm}
\rotatebox{270}{\resizebox{6.5cm}{!}{\includegraphics{M30-modes.eps}}}
\end{center}
\caption{Adiabatic pulsation frequencies of radial (blue circles), dipole
  (black crosses), and quadrupole (red triangles) zonal modes for a stellar model with
  $M=35\,$M$_\odot$ (left) and with $M=30\,$M$_\odot$ (right). EPIC\,202060092 is
  situated in between the two vertical dashed lines, which indicate the
  spectroscopic $1\sigma$-range of the measured $T_{\rm eff}$. The top $x-$axes indicate the
  central hydrogen fraction, where the value at birth was $X_c=0.715$. The
  horizontal grey lines indicate the five dominant independent frequencies, where
  the line thickness represents the frequency precision.}
\label{M35-30-modes}
\end{figure*}

The five dominant frequencies listed in Table\,\ref{star92} have values typical
of pressure modes in an O9 main-sequence star as already found for HD\,46149 and
HD\,46202 by \citet{Degroote2010} and \citet{Briquet2011}, respectively. Part of
the theoretical radial and zonal dipole or quadrupole p mode spectrum for O-type
stars is shown in Fig.\,\ref{M35-30-modes} for a mass of 35\,M$_\odot$ (left
panel) and 30\,M$_\odot$ (right panel). This figure was constructed from the
grid of stellar and pulsational models described in \citet{Degroote2010},
whose input physics is discussed in \citet{Briquet2011}, to which we refer
for details. The two stellar models whose zonal p modes are shown in
Fig.\,\ref{M35-30-modes} have an initial hydrogen fraction of $X_c=0.715$ and a
step-function core overshoot parameter of $\alpha_{\rm ov}=0.2$, expressed in
units of the local pressure scale height. As we have shown in the previous Section,
EPIC\,202060092 is situated between the two vertical dashed lines. These
correspond to ages of 3.53 and 3.51\,Myr (hotter temperature limit) and 4.39 and
4.86\,Myr (coolest temperature limit), for 35\,M$_\odot$ and 30\,M$_\odot$,
respectively.

Even though relatively large uncertainties occur for the
atmospheric parameters of the star, the measurement of $v\sin\,i$ is fairly
robust. It immediately excludes $f_{12}=0.1123\,$d$^{-1}$ to be the rotational frequency. Indeed,
the models with 35\,M$_\odot$ and 30\,M$_\odot$ whose low-degree p modes were
shown in Fig.\,\ref{M35-30-modes} have radii of 15\,R$_\odot$ and
12.5\,R$_\odot$, respectively, for the measured $T_{\rm eff}$. We find lower
limits for the rotational splitting within p mode multiplets of roughly
0.36\,d$^{-1}$ and 0.43\,d$^{-1}$ for 35\,M$_\odot$ and 30\,M$_\odot$,
respectively.

Even though the white-light CoRoT photometry of HD\,46202 by itself did not lead to mode
identification in the absence of multiplets in the frequency spectrum, the same situation as we have
here for EPIC\,202060092, \citet{Briquet2011} managed to perform forward seismic
modelling from frequency matching and pinpointed the four free parameters
$(M,X,Z,\alpha_{\rm ov})$ of HD\,46202 with high precision, for the
input physics that went into their dense grid of theoretical models. 
Their frequency precision was much higher, typically between $10^{-4}$ and
$10^{-3}$\,d$^{-1}$, than what we can deduce from the current long-cadence K2
data. This major accomplishment for HD\,46202
was owed to the CoRoT sampling rate of 32\,s,
which delivered a light curve of more than 80\,000 data points rather than only
1\,600 for EPIC\,202060092 from K2. Despite similar duration of the monitoring,
this implies a factor $\sim\sqrt{50}$ better frequency precision. Moreover,
their study was based on twice as many spectra of a single star of one magnitude
brighter than EPIC\,202060092, implying a co-added averaged spectrum of SNR
above 500.  Finally, the ten times lower $v\sin\,i=25\pm 7\,$km\,s$^{-1}$ of
HD\,46202 compared to EPIC\,202060092 also implied modest rotational splitting
values such that forward modelling assuming zonal modes made sense.  

For the
case of EPIC\,202060092, which has a similar oscillation spectrum and hence the
same potential, we encounter the limitations of the poor frequency precision and
of not knowing the azimuthal order $m$ of the dominant modes because the fast rotation
introduces frequency shifts of at least 0.3\,d$^{-1}$ when
$m\neq\,0$. It can be seen from Fig.\,\ref{M35-30-modes} that, even for only two
masses and one set of values $(X,Z,\alpha_{\rm ov})$, several of these models fulfill
the requirements of the five measured frequencies, keeping in mind that large
yet unknown frequency shifts due to rotational splitting must be allowed for.  A
similar situation of unknown rotational shifts was encountered in the NGC\,884
cluster modelling by \citet{Saesen2013}, but in that study some seismic
constraints could still be achieved thanks to the demand of an equal age and
metallicity of the stars in the cluster.

With the present K2 photometry and {\sc HERMES} spectroscopy, forward seismic
modelling is not yet possible for EPIC\,202060092. We do note that even the
identification of $(\ell,m)$ of just one of its modes would break most of the
degeneracy in the seismic grid. As an example, identification of $f_4$ as a
radial mode would imply it is the fundamental, while $f_3$ is then the third
(for 35\,M$_\odot$, left panel of Fig.\,\ref{M35-30-modes}) or fourth (for
30\,M$_\odot$, right panel of Fig.\,\ref{M35-30-modes}) radial overtone.  At
present, however, such inference remains speculative. Nevertheless, the models
imply that, if $f_1$ to $f_5$ are due to low-degree low-order modes, then the
mass of the star cannot be far below 30\,M$_\odot$.

Firm conclusions on the properties of the orbit of EPIC\,202060092 and on the
nature of its companion require additional and long-term time-resolved high SNR
spectroscopy. Given its potential for future seismic modelling, we plan to acquire such
data with the aim to pinpoint the orbit and characterise the two components with
high precision, as well as to detect and identify the dominant oscillation
mode(s).  This will require spectral disentangling and the application of
sophisticated mode identification methodology \citep[e.g.,][for an example of
another potentially interesting pulsating massive binary]{Tkachenko2014}.

\section{State of affairs of high-precision O-star space photometry}

\begin{table*}
  \caption{Overview of O-star variability deduced from high-cadence high-precision
    space photometry with the dominant causes of the variability listed, where
    IGW stands for ``Internal Gravity Waves'' \citep[e.g.][]{Rogers2013}. The
    duration of the     longest uninterrupted light curve is listed.}
\centering
\tabcolsep=4pt
\begin{tabular}{lccccccl}
\hline
\hline
HD or EPIC & V & SpT & Data Source & Duration&$v\sin\,i$&Variability & Reference to\\
 &  mag&  &  & of LC (d) & km\,s$^{-1}$ & Cause(s) & Space Photometry\\
\hline
HD\,46223	& 7.28 & O4V(f) &CoRoT&34.3 &100 &IGW		& \citet{Blomme2011} \\[3pt]
HD\,66811 & 2.24 & O4I(n)fp & SMEI & 236 & 400 & non-adiabatic g$^-$ mode? & \citet{HowarthStevens2014}\\[3pt]
HD\,46150	& 6.73 & O5V(f) &CoRoT&34.3 & 80 & IGW		& \citet{Blomme2011} \\[3pt]
HD\,24912 & 4.06 & O7IIInf &MOST& 30.0 & 200 & Rotational Modulation & \citet{Ramiaramanantsoa2014}\\[3pt]
EPIC\,202060098	& 15.1 & O7V &K2&33.1&?&Rotational Modulation& this work \\[3pt]
HD\,46966	& 6.87 & O8V	    &CoRoT&34.3 &50 &IGW		& \citet{Blomme2011} \\[3pt]
HD\,47129      & 6.06       & O8 III/I+O7.5III             &CoRoT&34.3 & 75 \& 300 & Binarity, Rotation & \citet{Mahy2011}\\
	&  & &	&&& Magnetic secondary & \\[3pt]
HD\,46149	& 7.61 & O8.5V((f))+B? &CoRoT&34.3 &30 \& ?	& Rotational Modulation & \citet{Degroote2010} \\
	&  & &	&&& \& Stochastic p modes & \\[3pt]
HD\,46202	& 8.19 & O9V(f)&CoRoT&34.3 &25	& Heat-driven p modes & \citet{Briquet2011} \\[3pt]
 HD\,44597 &  9.05	& O9V &K2&33.1&$\sim\,$30&IGW?& this work \\[3pt]
 HD\,256035& 9.21	& O9V:p+? &K2&33.1&270&Heat-driven modes& this work \\[3pt]
 HD\,255055& 9.40	& O9V:p(e) &K2&33.1&$\sim\,40$&IGW?& this work \\[3pt]
 HD\,149757 & 2.56 & O9.5Vnn & WIRE,MOST,SMEI & 232& 400 & Heat-driven p modes & \citet{Howarth2014}\\[3pt]
 EPIC\,202060097   & 13.7 & O9.5V &K2&33.1&?&Rotational Modulation& this work \\
\hline
\end{tabular}
\label{summary}
\end{table*}

The five O-type stars monitored by K2 so far confirm earlier findings based on
high-precision uninterrupted space photometry, i.e., the diversity of the
variable character of these stars is large.  Table\,\ref{summary} provides an
overview of all fourteen O-type stars that have been monitored with SMEI, MOST,
CoRoT, and K2 and their dominant cause of the photometric variability.  For
several of the stars, various causes of variability act simultaneously.  The
secondary of the binary HD\,47129 is the only object in the list known to host a
magnetic field and it is variable.  Nevertheless, rotational modulation 
connected with chemical and/or temperature spots  occurs for 4 of the 14 stars 
and is usually due to 
some level of magnetic activity even though the fields may be too weak 
or too complex to give detectable longitudinal field components.

From the viewpoint of seismic modelling with the aim to improve the input
physics of massive star models, one needs to secure high-precision frequency
values of several pulsation modes. While this is not impossible and has been
achieved for a rapidly-rotating pulsating Be star \citep[e.g.][]{Neiner2012}, it
turned out hard to achieve so far in the case of internal gravity waves driven
by the convective core in O stars \citep[e.g.][]{AertsRogers2015}.  The
requirement to detect oscillation frequencies with high precision of
$\sim\,0.0001\,$d$^{-1}$ can best be achieved in the case of self-driven modes
caused by an opacity mechanism active in the partial ionisation layers of the
iron-group elements, because they have long lifetime.  HD\,46202 is the only
O-type star that has been modelled seismically so far \citep{Briquet2011}. While
this led to the derivation of its fundamental parameters with higher precision
than any other method delivered so far, it was not possible to probe its
interior structure properties at a level necessary to improve the input physics
of the models, due to the too short time base of the space data.

Heat-driven O-type pulsators hold the same potential of deep seismic sounding of
their interior structure than was recently achieved by \citet{Moravveji2015} for
the B8.3V star KIC\,10526294. This 3.2\,M$_\odot$ star provided seismic evidence
for diffusive mixing in the stellar envelope at a level of $\log\,D_{\rm mix}$
between 1.75 and 2.00\,dex, in addition to core overshooting.  A major improvement
for evolutionary models of massive stars requires the observational calibration
of the overall mixing properties in their interior, because theoretical
considerations lead to $\log\,D_{\rm mix}$ values differing by many orders of
magnitude and are therefore of limited value for the moment
\citep[e.g.,][]{Mathis2004}. Asteroseismology of O star pulsators is the best
method to make progress for the tuning of models of the most massive stars in
the Universe, but it requires long-term high-precision time-resolved
photometric and spectroscopic monitoring. As we have shown in this work, the K2
mission has a major role to play here and EPIC\,202060092 was an optimal target
among five measured ones, but its complex beating pattern was not sufficiently
sampled in terms of cadence, precision, and duration of the photometry achieved
during Campaign\,0. Moreover, it turned out to be a spectroscopic binary and we
need to understand the contributions of each of its components to the detected
variability.  Requirements for its future seismic modelling are the accurate
determination of its orbital properties and of the fundamental parameters of its
binary components, along with a higher precision of the pulsation
frequencies. Given its potential, spectroscopic monitoring will be continued
in the coming years with the aim to unravel the orbital motion and to attempt
detection of its oscillation modes in long-term residual spectroscopy.

Following \citet{Degroote2010} and \citet{Briquet2011}, our current work is a 
successful proof-of-concept
study to perform future O-star seismology. {\it This is possible with the optimised K2 mission in
combination with ground-based spectroscopy}.  Indeed, as of K2's Campaign\,3, the
improvement on its performance in terms of pointing and precision of the
photometry, as well as in the duration of the campaigns ($\sim\,80$\,d, i.e.,
twice to three times as long as the current study) holds the potential to
deliver the appropriate photometric data needed for in-depth modelling of O-star
pulsations, whatever their excitation mechanism.  A light curve at short cadence
during 80\,d will deliver some 115\,000 data points, which will result in a
factor $\sim\,20$ better frequency precision than for the Campaign\,0 data.
Given their scarcity and importance for stellar and galactic structure, we thus
plan to apply for short-cadence data for the few available O stars in K2's future
FoVs of Campaigns 11 and 13.

\section*{Acknowledgments}
Part of the research included in this manuscript was based on funding from the
Research Council of KU\,Leuven, Belgium under grant GOA/2013/012, from the
European Community's Seventh Framework Programme FP7-SPACE-2011-1, project
number 312844 (SpaceInn), and from the Fund for Scientific Research of Flanders
(FWO) under grant agreement G.0B69.13. Funding for the Kepler mission is
provided by the NASA Science Mission directorate.  The authors wish to thank the
entire {\it Kepler\/} team and the {\it Kepler\/} Guest Observer Office for all
their efforts.

\appendix

\section{Masks for light curve extraction}

\begin{figure*}
\begin{center}
\rotatebox{0}{\resizebox{8.5cm}{!}{\includegraphics{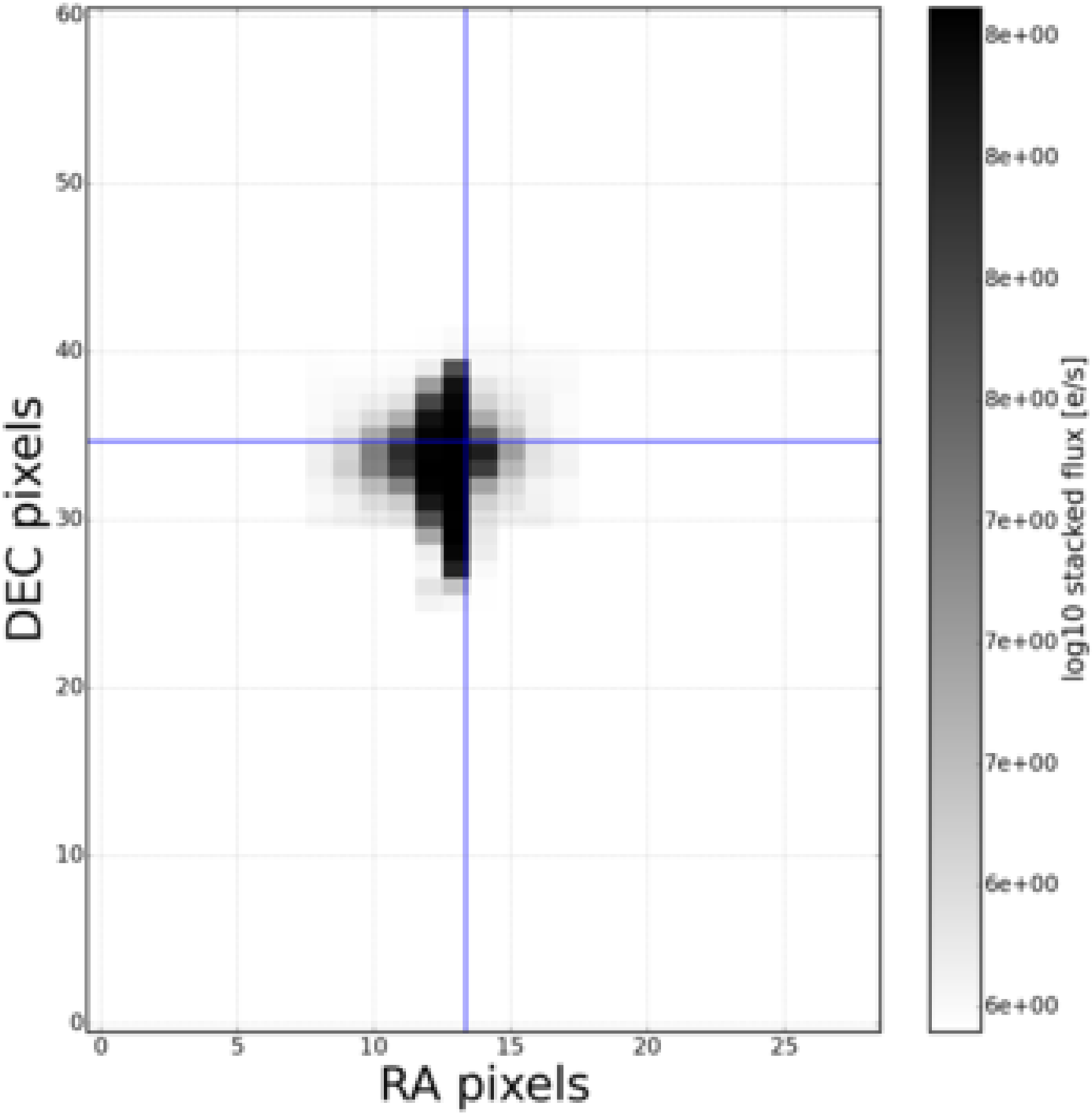}}}\hspace{0.4cm}
\rotatebox{0}{\resizebox{8.5cm}{!}{\includegraphics{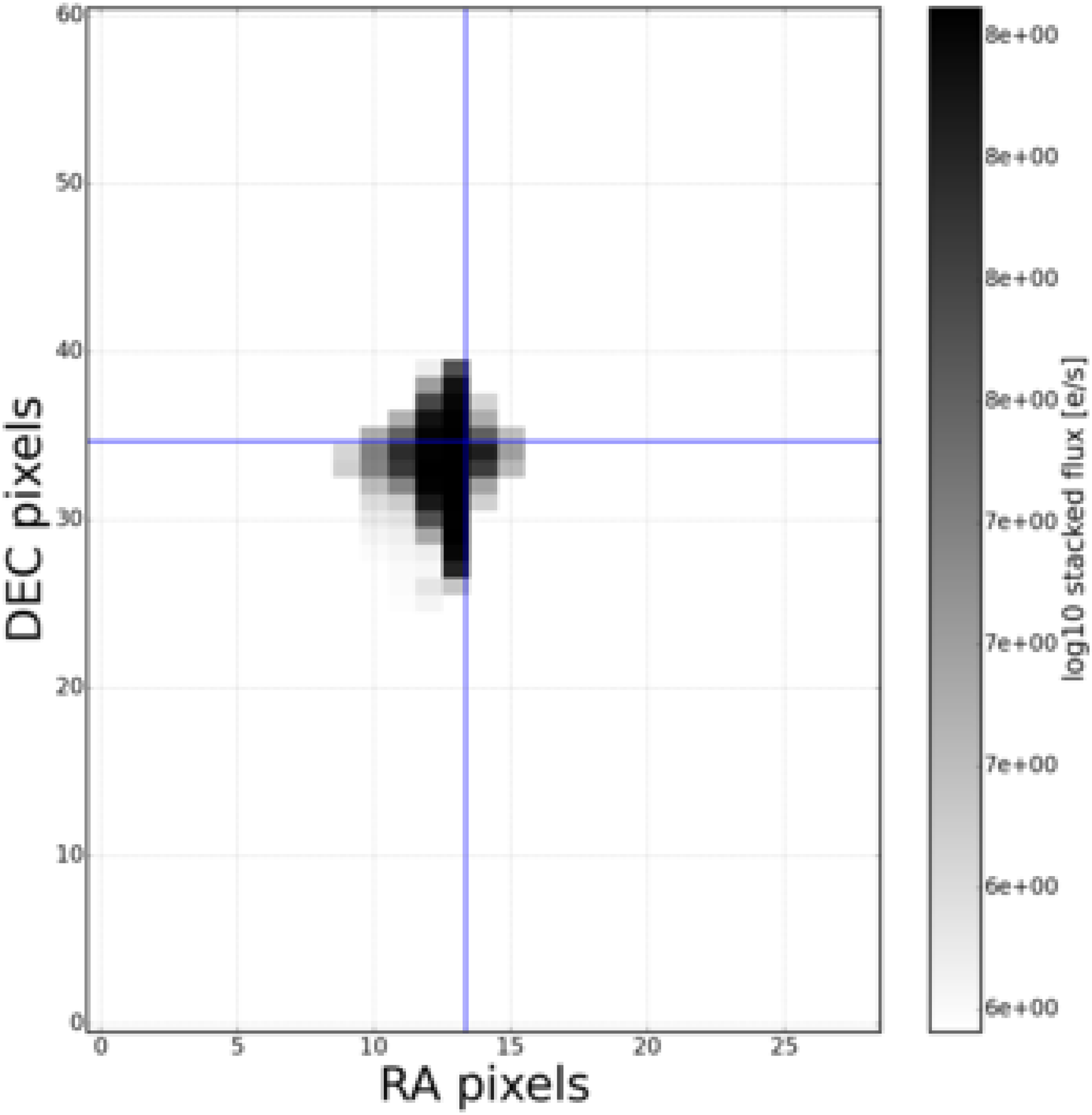}}}
\end{center}
\caption{Our adopted mask (left) versus the one used by Vanderburg (right)
for  the light curve extraction of EPIC\,202060091 as indicated in pink. The
position of the star is indicated by the blue cross.}
\label{91-mask}
\end{figure*}

\begin{figure*}
\begin{center}
\rotatebox{0}{\resizebox{8.5cm}{!}{\includegraphics{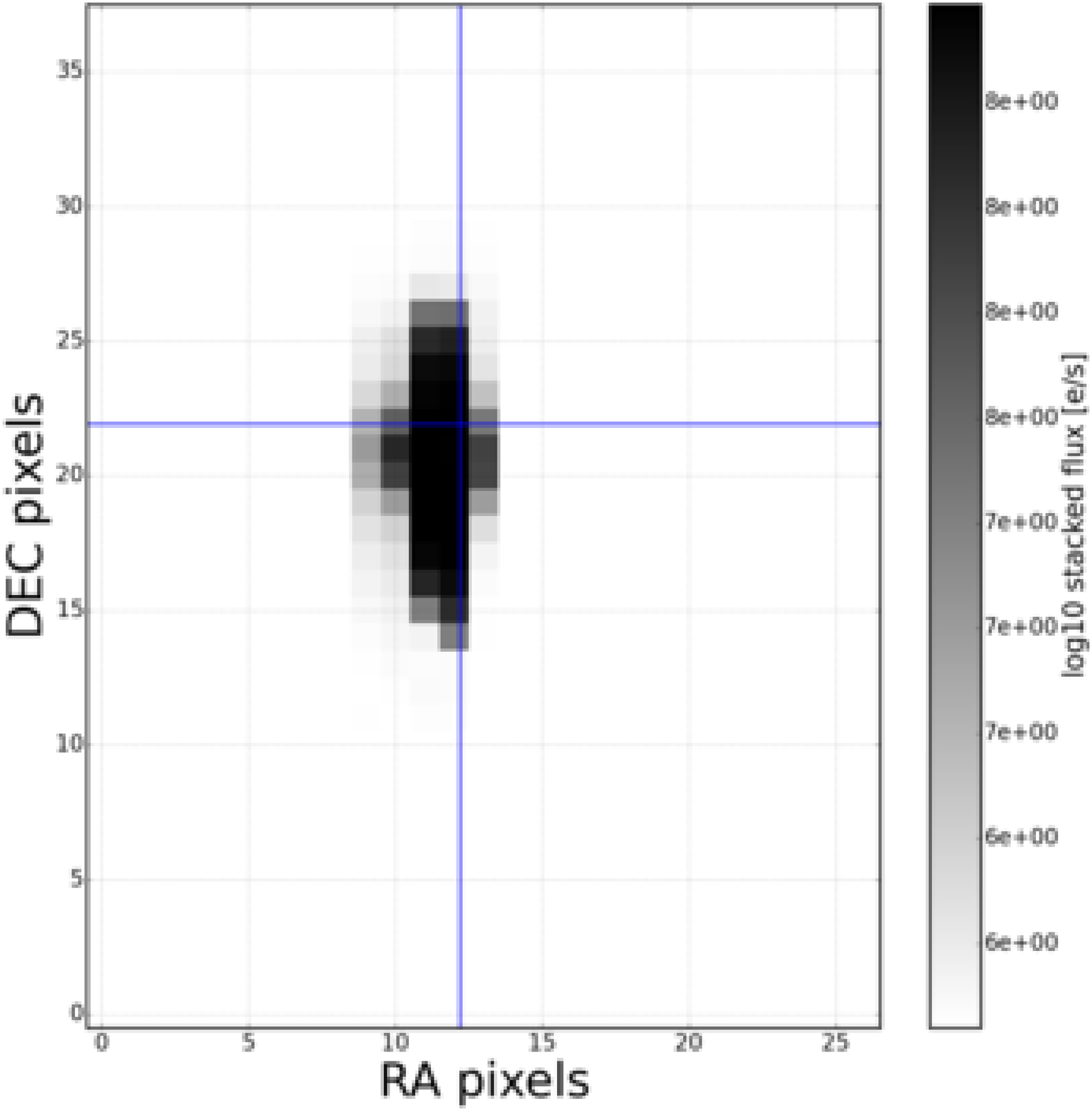}}}\hspace{0.4cm}
\rotatebox{0}{\resizebox{8.5cm}{!}{\includegraphics{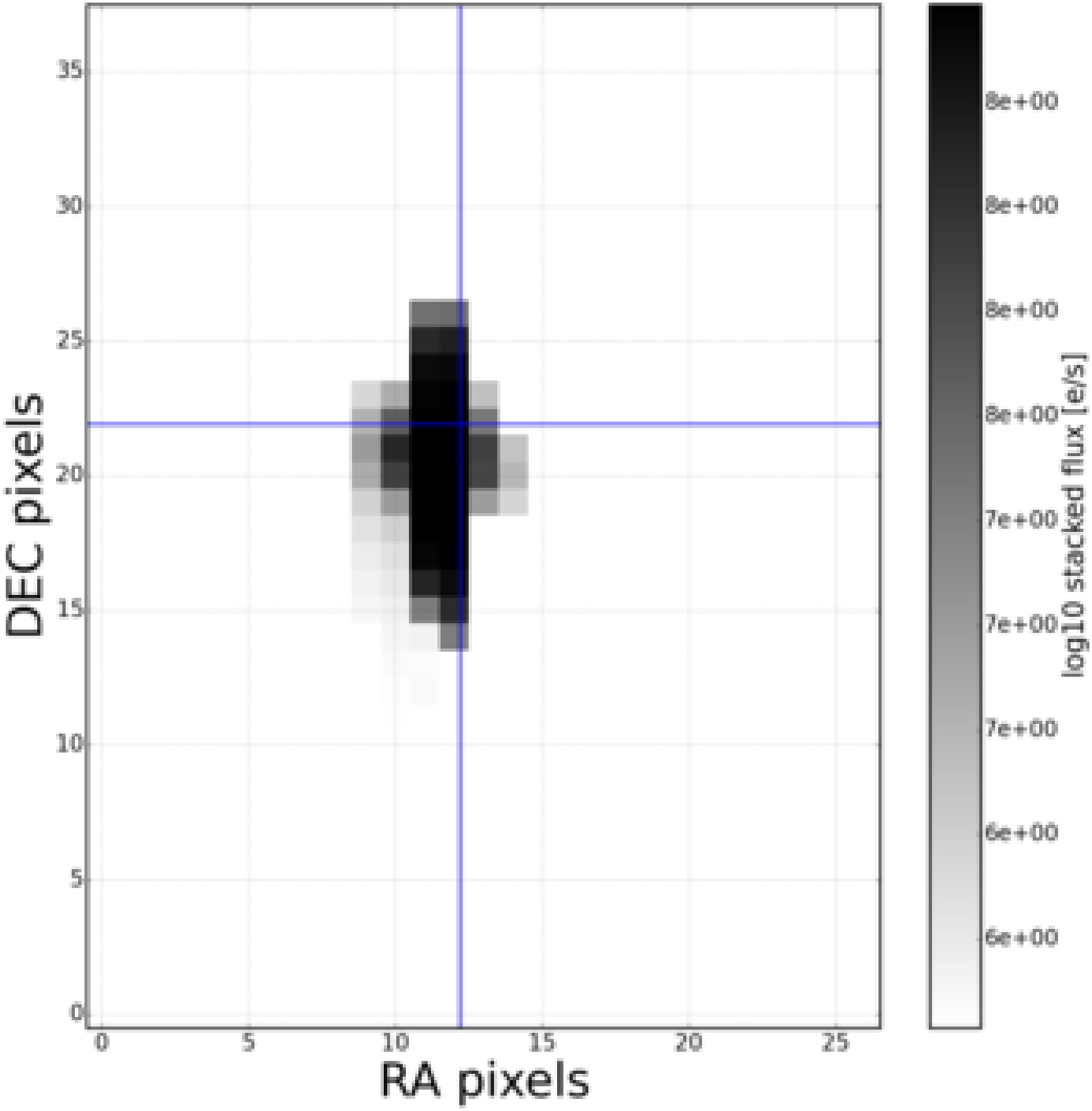}}}
\end{center}
\caption{Same as Fig.\,\ref{91-mask}, but for EPIC\,202060092.}
\label{92-mask}
\end{figure*}

\begin{figure*}
\begin{center}
\rotatebox{0}{\resizebox{8.5cm}{!}{\includegraphics{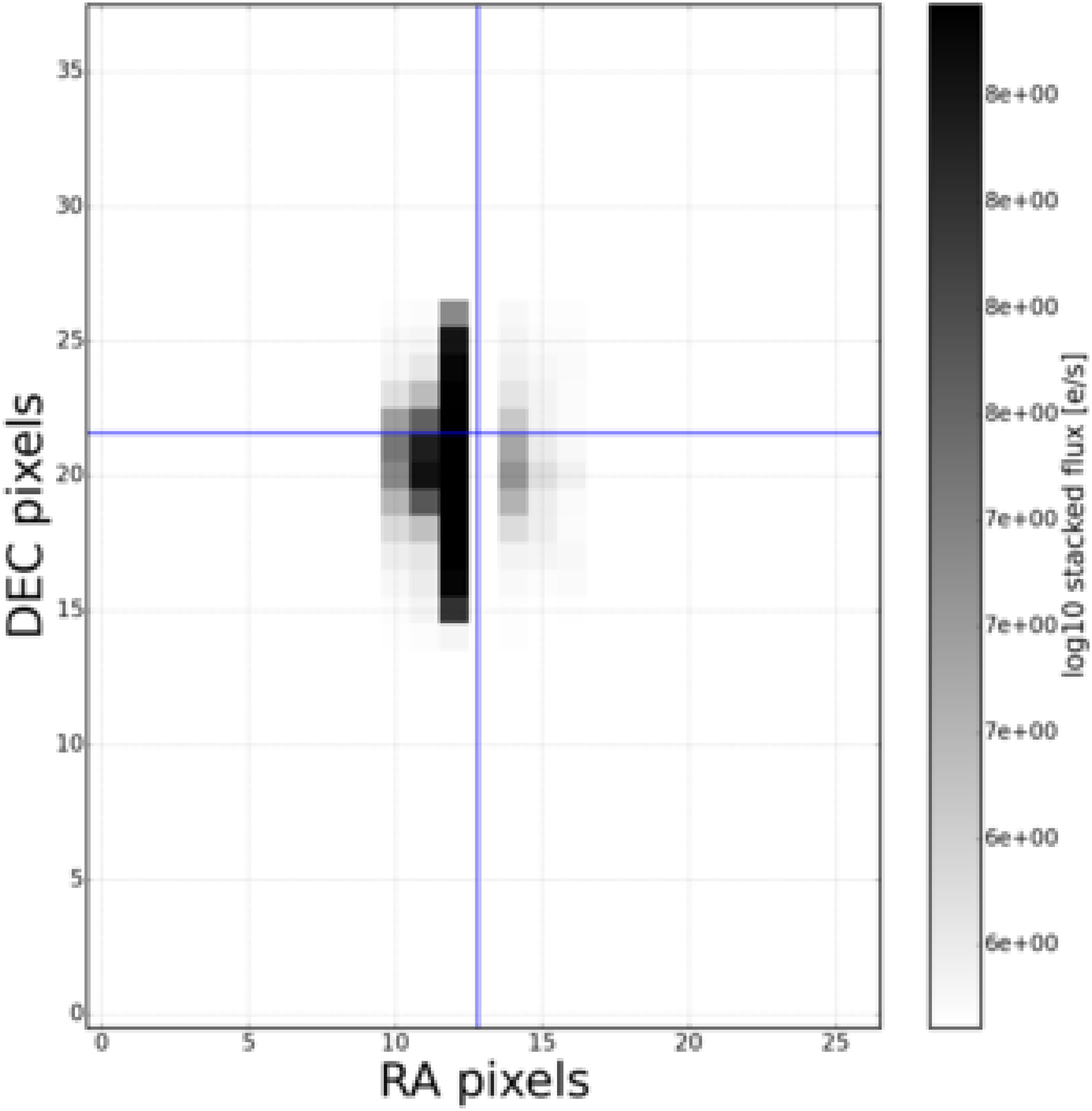}}}\hspace{0.4cm}
\rotatebox{0}{\resizebox{8.5cm}{!}{\includegraphics{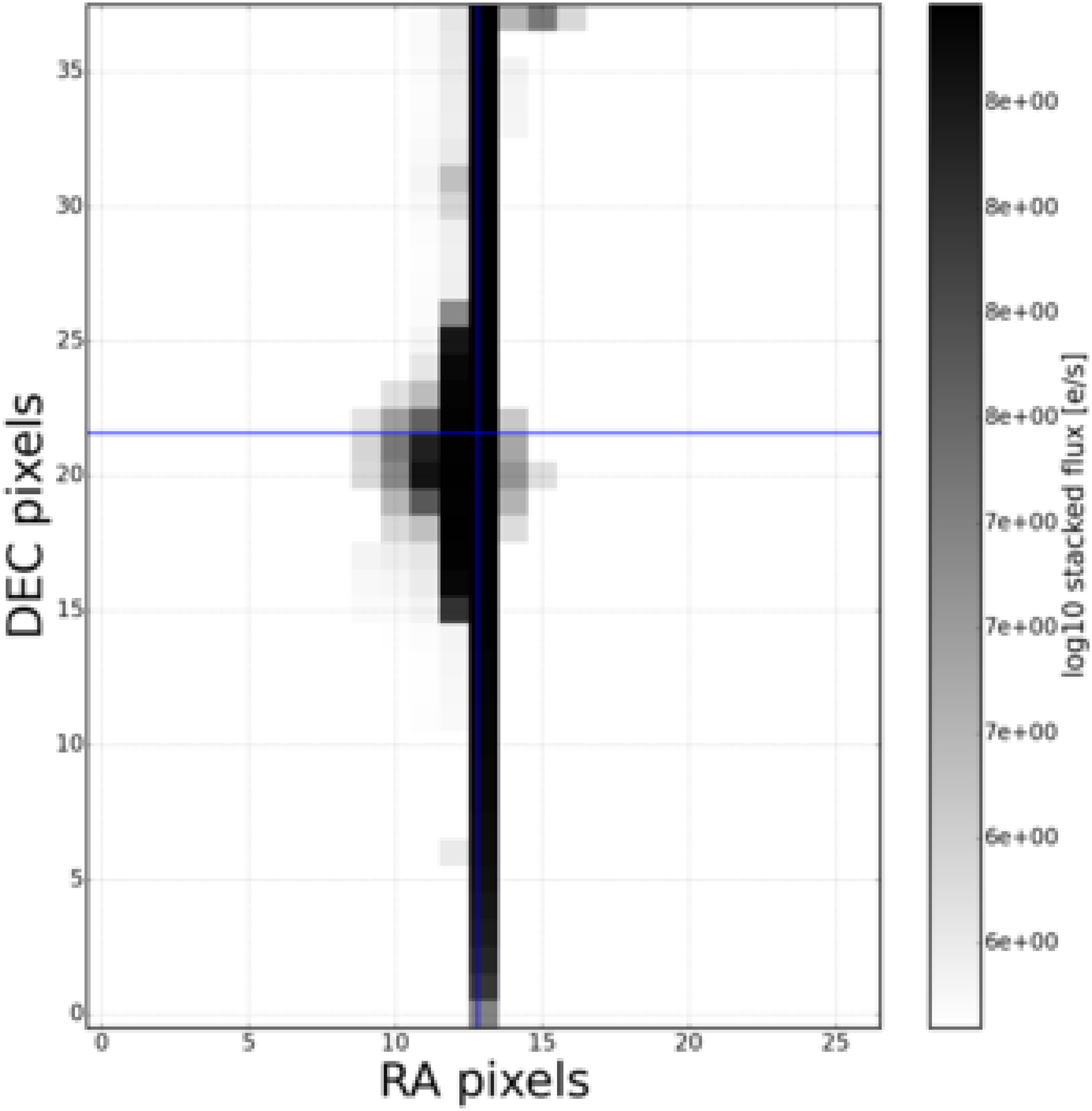}}}
\end{center}
\caption{Same as Fig.\,\ref{91-mask}, but for EPIC\,202060093. For this target,
  the mask on the left excluded the central saturated column to produce the
  photometry shown as black circles in the top panel of Fig.\,\ref{LC},
  while this column was used to produce the Vanderburg version of the light
  curve shown as grey crosses in the top panel of Fig.\,\ref{LC}.}
\label{93-mask}
\end{figure*}

\begin{figure*}
\begin{center}
\rotatebox{0}{\resizebox{8.5cm}{!}{\includegraphics{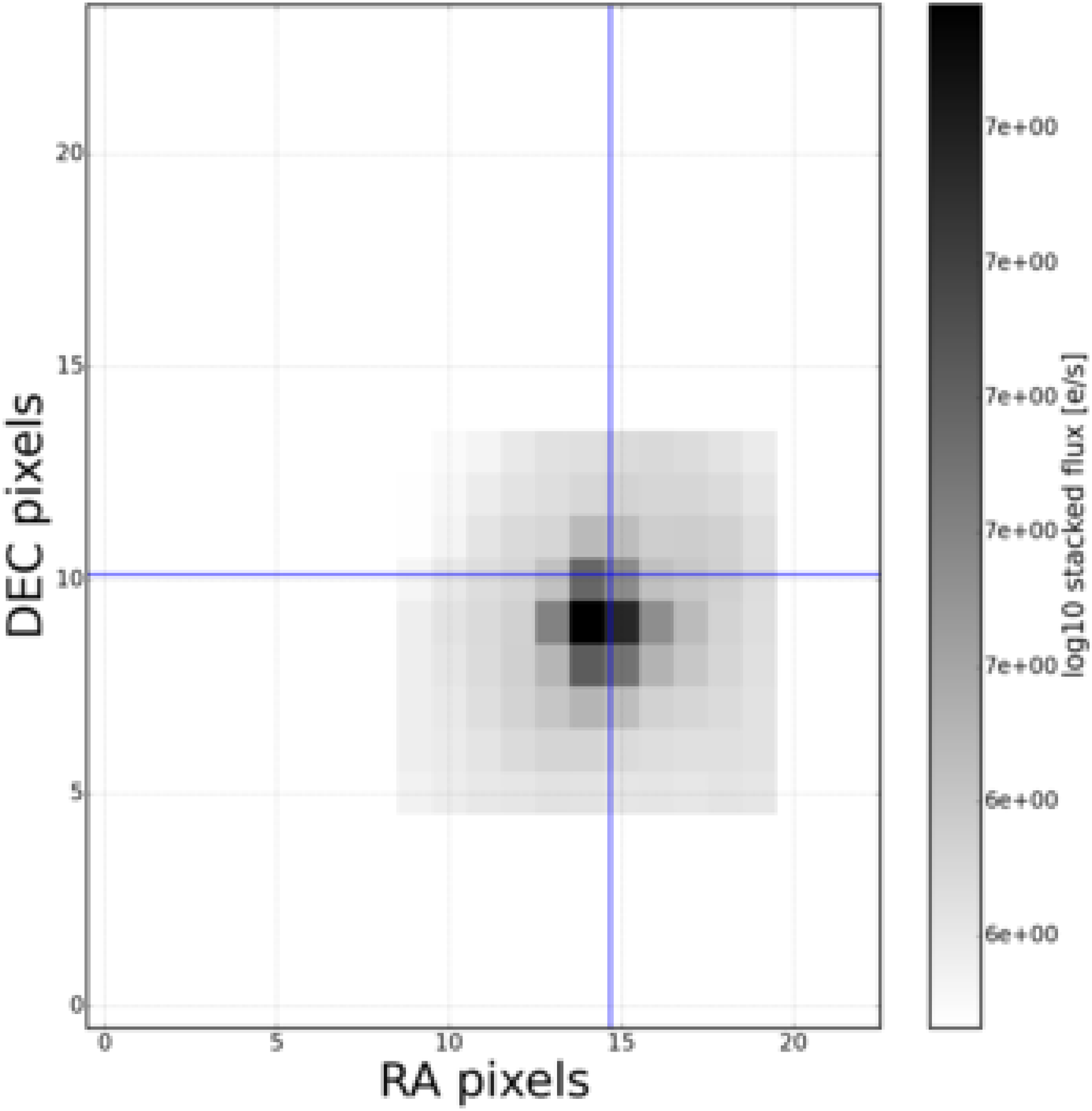}}}\hspace{0.4cm}
\rotatebox{0}{\resizebox{8.5cm}{!}{\includegraphics{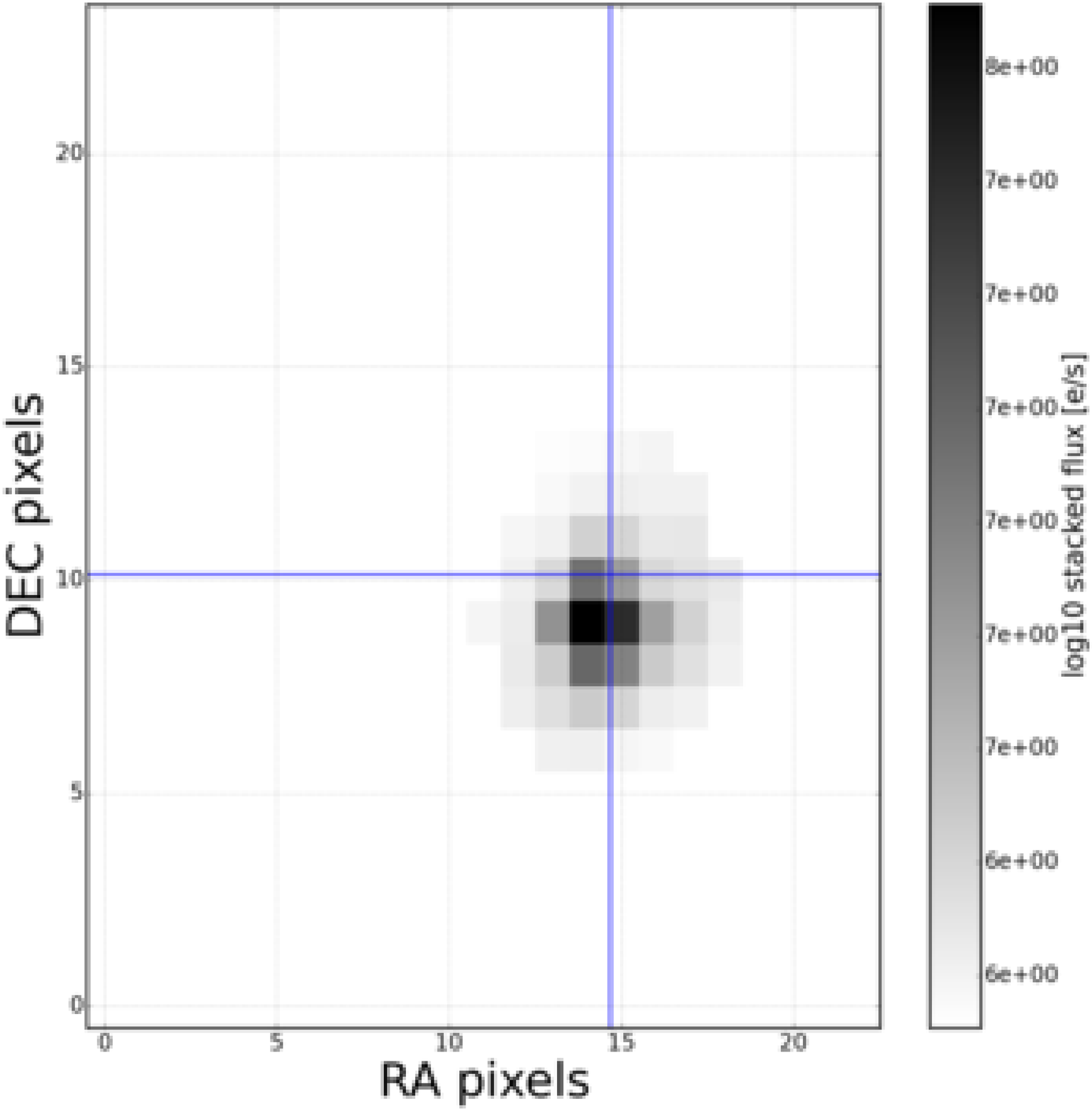}}}
\end{center}
\caption{Same as Fig.\,\ref{91-mask}, but for EPIC\,202060097.}
\label{97-mask}
\end{figure*}

\begin{figure*}
\begin{center}
\rotatebox{0}{\resizebox{8.5cm}{!}{\includegraphics{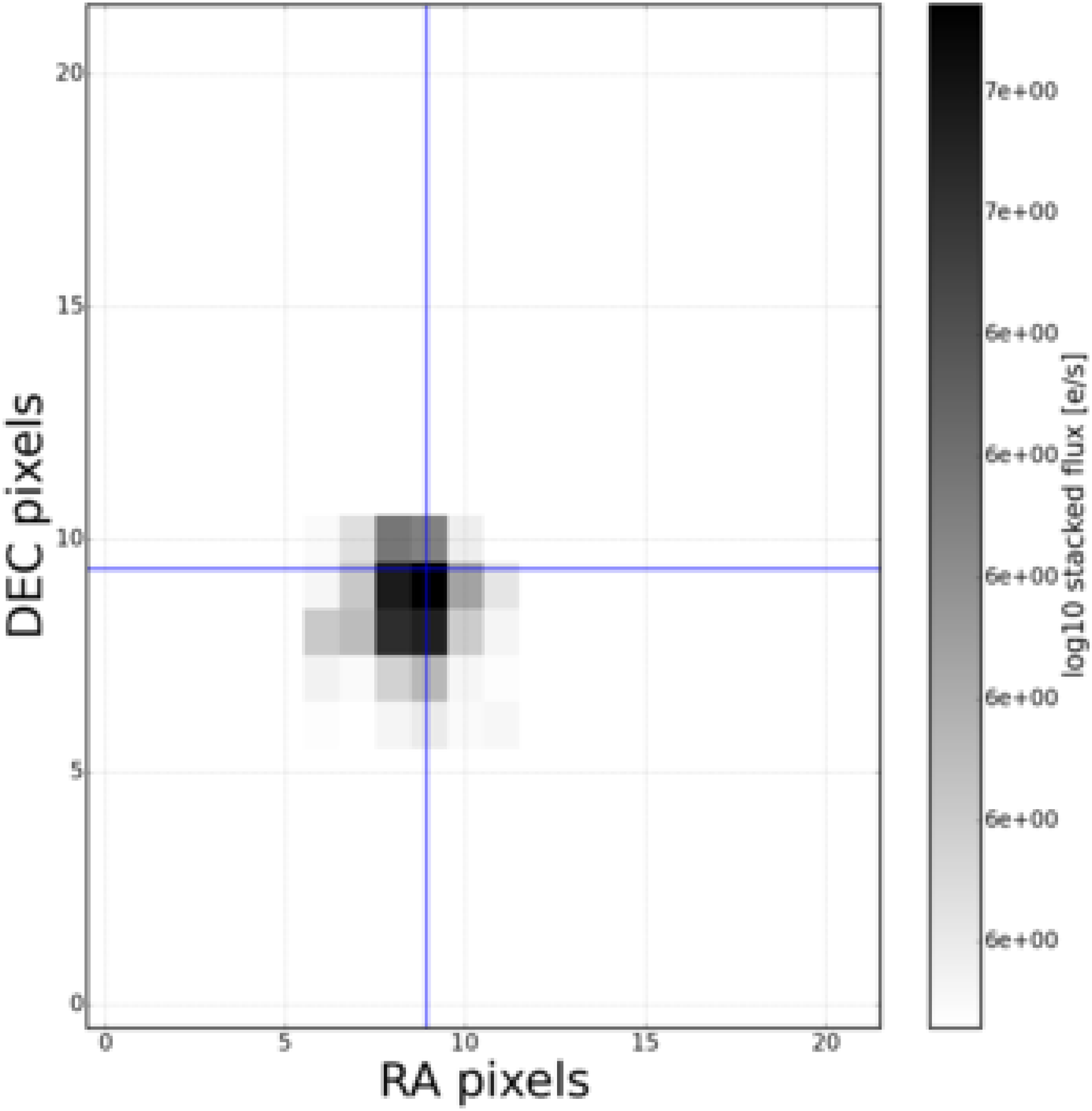}}}\hspace{0.4cm}
\rotatebox{0}{\resizebox{8.5cm}{!}{\includegraphics{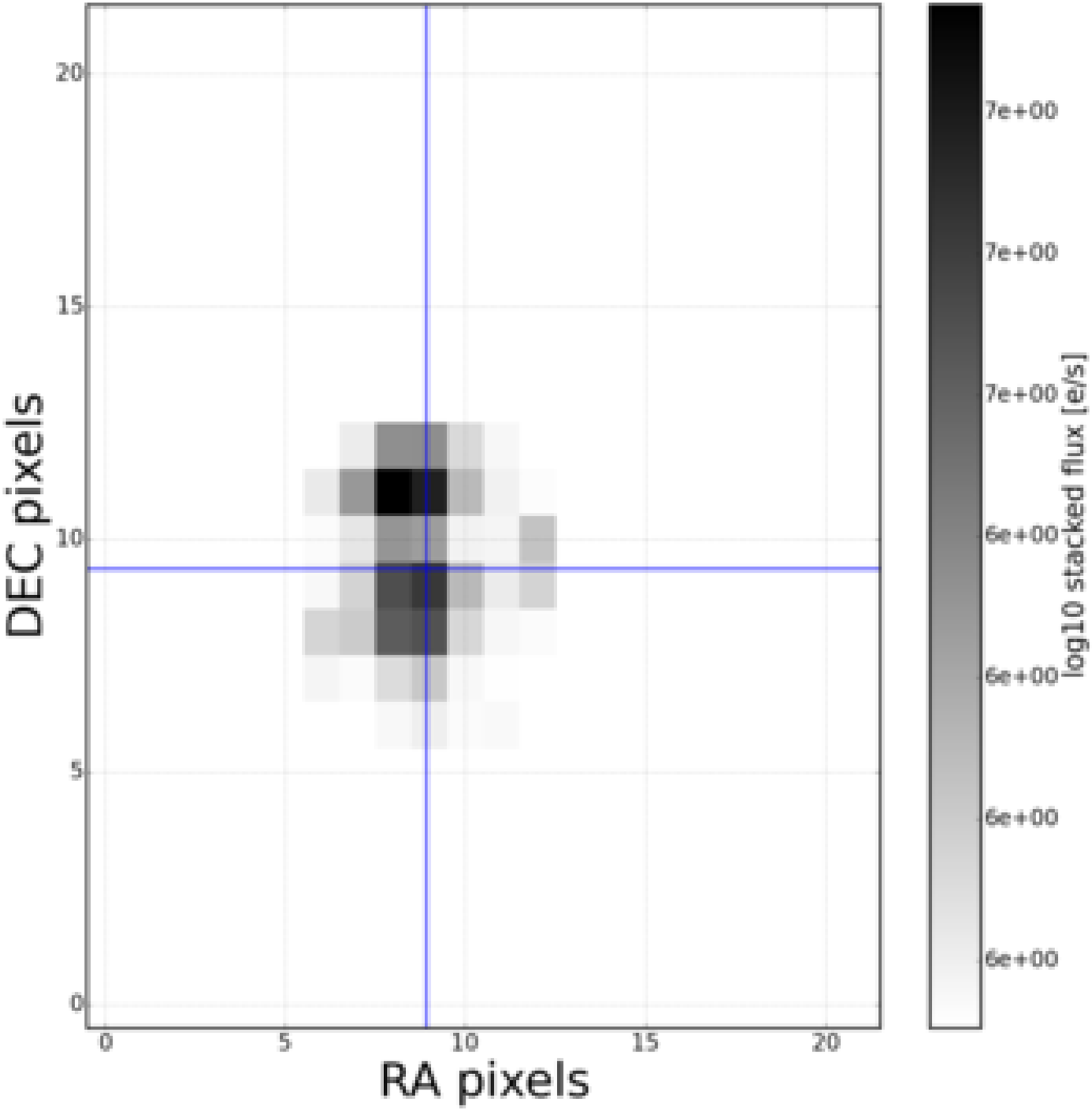}}}
\end{center}
\caption{Same as Fig.\,\ref{91-mask}, but for EPIC\,202060098.}
\label{98-mask}
\end{figure*}

\label{lastpage}

\end{document}